\documentclass[twoside,a4paper,11pt]{sca}
\usepackage{graphicx}
\usepackage{hyperref}
\usepackage{natbib}  
\usepackage{amssymb} 

\newcommand{\strom}{\mbox{Str\"omgren~}}

\def\bmy{\hbox{\it b--y\/}}
\def\vmy{\hbox{\it v--y\/}}
\def\camy{\hbox{\it Ca--y\/}}
\def\umy{\hbox{\it u--y\/}}
\def\feh{\hbox{\rm [Fe/H]}}
\def\mh{\hbox{\rm [M/H]}}
\def\afe{\hbox{\rm [$\alpha$/Fe]}}

\begin{document}
\pagenumbering{arabic}
\pagestyle{myheadings}
\thispagestyle{empty}
{\flushright\includegraphics[width=\textwidth,bb=90 650 520 700]{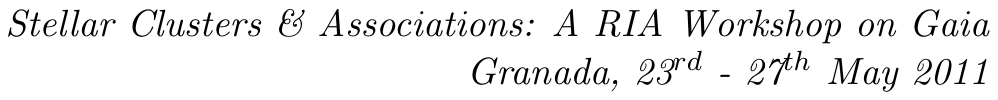}}
\vspace*{0.2cm}
\begin{flushleft}
{\bf {\LARGE
%
Str\"omgren photometry of the Bulge: the Baade's Window and the globular cluster
NGC~6522 
%
}\\
\vspace*{1cm}
%
A. Calamida$^{1}$,
G. Bono$^{2}$, 
C. E. Corsi$^{1}$,
G. Iannicola$^{1}$,
V. Ripepi$^{3}$,
B. Anthony-Twarog$^{4}$,
B. Twarog$^{4}$,
M. Zoccali$^{5}$,
R. Buonanno$^{2}$,
S. Cassisi$^{6}$,
I. Ferraro$^{1}$,
F. Grundahl$^{7}$,
A. Pietrinferni$^{6}$,
and
L. Pulone$^{1}$
%
}\\
\vspace*{0.5cm}
%
$^{1}$
Osservatorio Astronomico di Roma/INAF\\
$^{2}$
Universit\`a di Roma Tor Vergata\\
$^{3}$
Osservatorio Astronomico di Capodimonte/INAF\\
$^{4}$
University of Kansas\\
$^{5}$
Pontificia Universidad Cat\'olica de Chile\\
$^{6}$
Osservatorio Astronomico di Collurania/INAF\\
$^{7}$
Department of Physics and Astronomy, Aarhus University
\end{flushleft}
%
\markboth{
Str\"omgren photometry of the Bulge
}{ 
Calamida et al.
%
}
\thispagestyle{empty}
\vspace*{0.4cm}
\begin{minipage}[l]{0.09\textwidth}
\ 
\end{minipage}
\begin{minipage}[r]{0.9\textwidth}
\vspace{1cm}
\section*{Abstract}{\small
We present $Ca-uvby$ Str\"omgren~ photometry of the Baade's Window, 
including the Galactic globular cluster NGC~6522. 
We separate field and cluster stars by adopting color--color planes 
and proper motions. We then estimate the global metallicity of 
red-giants (RGs) in NGC~6522 by using a new theoretical metallicity 
calibration of the Str\"omgren index $hk$ presented in \citep{calamida11}.
We find that metallicities estimated by adopting the $hk$ index and the 
$\camy$ color are systematically more metal-rich than metallicities 
estimated with $hk$ and the $\umy, \vmy$ and $\bmy$ colors. 
Current evidence indicate that the $hk$ metallicity index is affected 
not only by the $Ca$ abundance, but also by another source of opacity. 
\normalsize}
\end{minipage}
%
%
%

\section{Introduction and data reduction}\label{obs_cala}
The study of the Milky Way bulge is fundamental to understand the
star formation history of the Galactic spheroid. 
The Milky Way bulge is indeed dominated by metal-rich old stars 
($t > 10$ Gyr, \citealt{zocc03}).
Their metallicity distribution is centered on solar metallicity 
and span a range from $[Fe/H] \sim$ -1.5 to 0.5 dex, based on 
the high-resolution spectroscopy of $\sim$ 800 stars 
\citep[hereafter ZO08]{zocc08}.
These evidence would support a scenario in which the bulge formed very quickly
($t <$ 1 Gyr) at the early stages of the Galaxy formation, and 
then accreted material from the disk (\citealt{zocc06}, ZO08).

Furthermore, the Baade's window, where the extinction is 
lower ($E(B-V) <$ 0.6 mag), includes NGC~6522, which is classified as
a bulge metal-intermediate Galactic globular cluster (GGC, $[Fe/H] \sim$ -1.3/-1.2 dex). 
For this region ZO08 collected high-resolution spectroscoy for $\sim$ 200 
red-giants (RGs) and $\sim$ 200 red clump stars. The two samples show the same
metallicity distribution centered on solar metallicity and with most of the stars
in the range -1.0 $< [Fe/H] <$ 0.7 dex (see Fig.~8 in ZO08).

We collected a set of {\it Ca-uvby\/} \strom images centered on the Baade's 
and the Blanco's windows and we plan to provide metallicity distributions 
based on the $hk$ Str\"omgren~ index for a significant sample of bulge RGs.
The advantage of applying a \strom metallicity calibration to this data 
set is the possibility to estimate the metallicity of thousands of RGs in 
this region at the same time. 
With this study we would be able to eventually confirm the presence of 
a metallicity gradient in the bulge, as suggested by ZO08.
The drawback is that our new theoretical metallicity calibration of the
$hk$ index (\citealt{calamida11}, hereafter CA11) 
is only valid in the range -2.7$< [Fe/H] <$-0.6 dex.
We then apply the calibration to a significant sample of NGC~6522 
candidate RGs, to constrain the metal content of this GGC.

\begin{figure}
\center
\includegraphics[width=15truecm,height=9truecm]{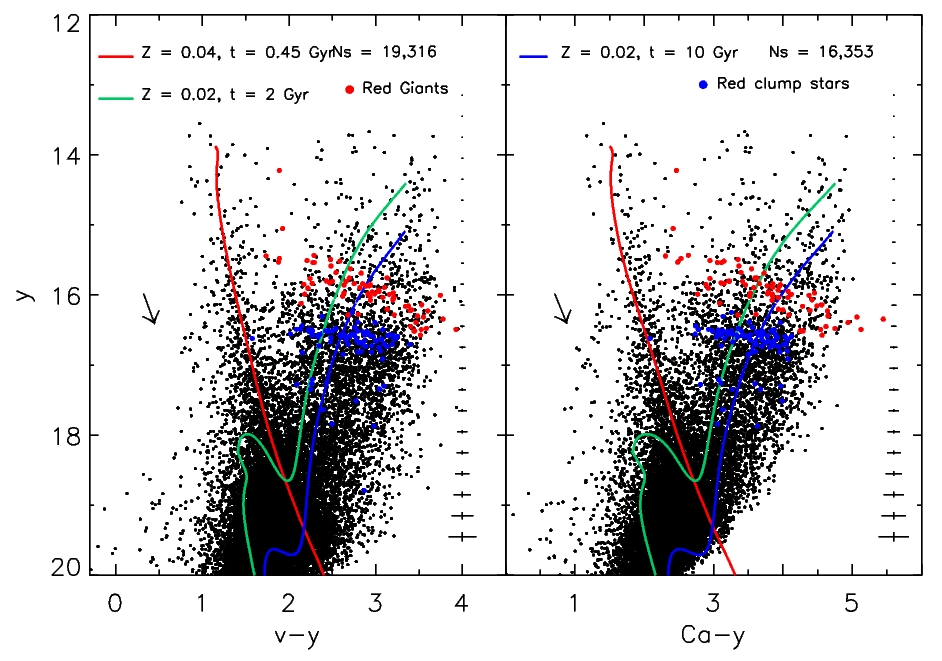} 
\vspace{-0.2cm}
\caption{\label{fig1} $y$, \vmy\ and 
$y$, \camy\ CMDs for stars in the Baade's Window.
Stars are selected in position and in photometric accuracy (see 
text for more details).
Error bars display intrinsic errors in color and in magnitude, 
while the arrows show the reddening directions. 
The red solid line shows an isochrones for $t = 0.45$ Gyr and $Z=0.04, Y=0.303$, 
while the green and blue solid lines show two isochrones for the same chemical
composition, $Z=0.02, Y=0.273$, and $t = 2$ and $t = 10$ Gyr, respectively.
Red and blue dots mark red-giant and red-clump stars observed 
spectroscopically by \citet{zocc08}.}
\end{figure}

{\it Ca-uvby\/} \strom images were collected with the 1.54m Danish Telescope 
(ESO, La Silla) and the DFOSC camera in July $2000$. 
The pointing was centered on the Baade's Window 
($\alpha$ = 18:03:34, $\delta$ = -30:04:10), including  NGC~6522. 
We secured a total of 16 images ($4y,4b,2v,2Ca,2u$), 
with exposure times ranging from 60s ($y$) to 1000s ($Ca$), and seeing 
between $\sim$1.2" and $\sim$1.6". 
The reader interested in the details of data
reduction and calibration is referred to CA11.
The final photometric catalog includes $\sim$80,000 stars with an 
accuracy of $\sigma_y <$ 0.1, $\sigma_{v-y} <$ 0.2 mag 
at $y \approx$ 20 mag. The accuracy of the calibration is $\sim$ 0.02 mag for 
the $y,b,v$ bands and $\sim$ 0.05 mag for the $Ca,u$ bands.

Fig.~\ref{fig1} shows the $y$, \vmy\ (left panel) and 
$y$, \camy\ (right) Color-Magnitude Diagrams (CMDs) of the Baade's Window.
We exclude most of the stars belonging to NGC~6522 by selecting only
objects with distances from the cluster center 
($\alpha$ = 18:03:34, $\delta$ = -30:02:02) larger than 6 arcminutes.
Stars are then selected in photometric accuracy according to the 
''separation index''\footnote{The ''separation index'' quantifies the degree 
of crowding \citep{ste03}.}

In order to validate the absolute calibration we compare our photometry 
with theoretical predictions.
We assume a mean distance of $\sim$ 5 Kpc for the foreground thin disk stars 
according to the simulations by Schultheis and using the Besancon Galaxy model 
(\citealt{robin03}, see ZO08 for more details), and a distance modulus of 
$\mu_0=13.91$ and a mean reddening of $E(B-V)=0.55$ 
for bulge stars \citep{bar98}. The extinction coefficients for 
the \strom colors are estimated by applying the \citet{card89} 
reddening relation and $R_V = A_V/E(B-V) = 3.13$ \citep{bar98}.
We find:  
$E(b-y)= 0.69\times E(B-V)$, 
$E(v-y)= 1.31\times E(B-V)$, 
$E(u-y)= 1.82\times E(B-V)$ and
$E(Ca-y)= 1.46\times E(B-V)$.

The red solid line shows an isochrones for $t = 0.45$ Gyr and 
$Z=0.04, Y=0.303$, while the green and blue solid lines show two 
isochrones for the same chemical composition, $Z=0.02, Y=0.273$, and
different ages, $t = 2$ Gyr and $t = 10$ Gyr, respectively.
Red and blue dots show RG and red-clump stars observed 
spectroscopically by ZO08. Isochrones are from 
the BASTI data base and are based on $\alpha$-enhanced ($\afe=0.4$) 
evolutionary models \citep{pietri06}. 
Evolutionary prescriptions were transformed into the observational 
plane using atmosphere models computed assuming $\alpha$-enhanced mixtures. 
Data plotted in Fig.~\ref{fig1} indicate that theory 
and observations, within the errors, agree quite well both with 
the thin disk main sequence (1.0 $\lesssim (v-y) \lesssim $ 2.5 mag, 
14 $\lesssim y \lesssim$ 20 mag, red solid line) and the bulge
and thick disk sequences (2.0 $\lesssim (v-y) \lesssim $ 3.5 mag, 
14 $\lesssim y \lesssim$ 18 mag, green and blue solid lines).
The spread of the sequences is mostly due to photometric errors and 
to the presence of differential reddening, but also to depth 
differences of the stars.

\begin{figure}
\center
\includegraphics[width=15truecm,height=9truecm]{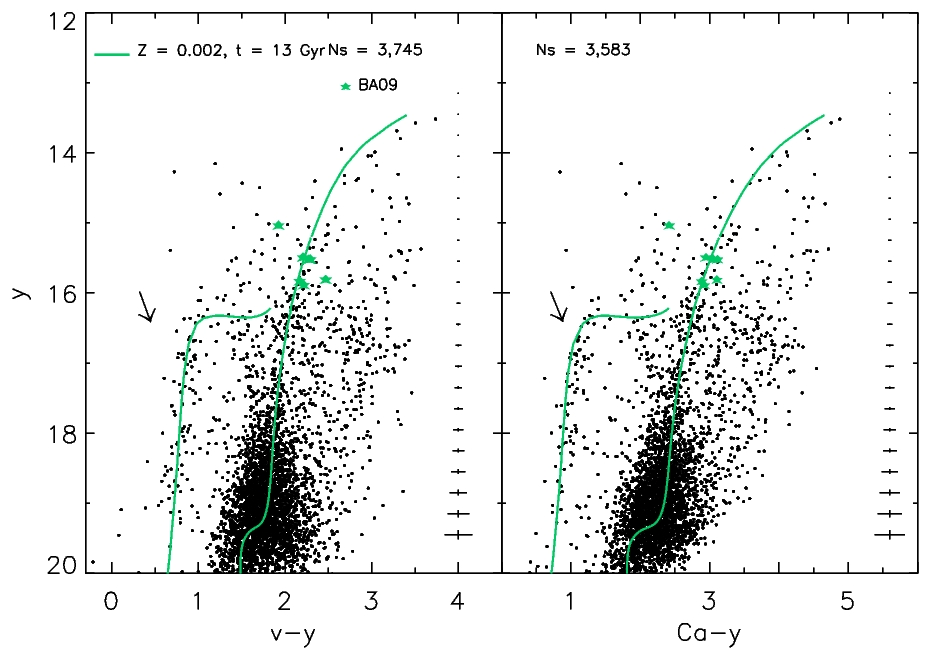} 
\vspace{-0.2cm}
\caption{\label{fig2} Same as Fig.~\ref{fig1} but for NGC~6522. 
The green solid lines show an isochrone for $t = 13$ Gyr and $Z=0.002, Y=0.248$ 
and the predicted ZAHB for $Z=0.002$, while the green stars mark the 
eight cluster red-giants observed spectroscopically by \citet{bar09}.}
\end{figure}

Fig.~\ref{fig2} shows the same CMDs but for NGC~6522: only stars 
with distances from the cluster center in the range 0.65 - 1.65 arcminutes 
are plotted. The cluster center is excluded 
due to crowding, but stars up to about 1.5 the half-light radius 
($r_h = 1.0$', \citealt{harris}) are selected (CA11).
The green solid lines show an isochrone for $t = 13$ Gyr
and $Z=0.002, Y=0.248$ and the predicted Zero Age Horizontal Branch 
(ZAHB) for $Z=0.002$.
The eight green stars are RGs with high-resolution spectra 
collected with FLAMES/GIRAFFE at the VLT (ESO, \citealt[hereafter BA09]{bar09}).
Note that theory and observations agree quite well and give \feh $\sim$ -1.3 dex for 
NGC~6522, in agreement with the spectroscopic estimate \feh $\sim$ -1.2 dex by BA09, 
converted to the \citet{zinn} metallicity scale (CA11).
\begin{figure}  
\center
\includegraphics[width=15truecm,height=9truecm]{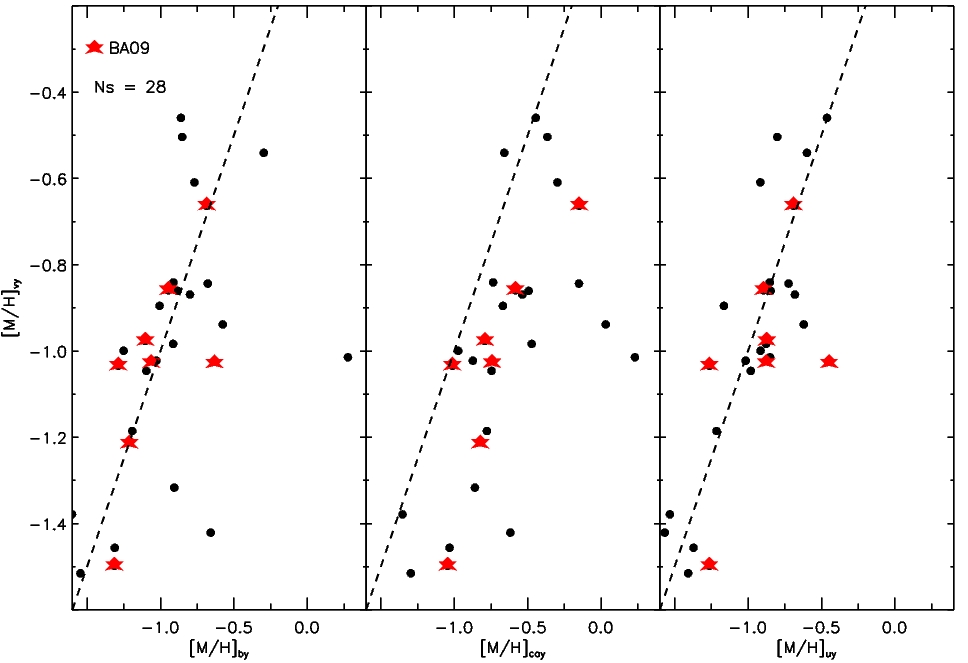}
\vspace{-0.2truecm}
\caption{\label{fig3} Photometric metallicities of 28 candidate 
RGs of NGC~6522 estimated adopting the Metallicity--Index--Color relations $hk_0,\ \bmy_0,\ \camy_0,\ \umy_0$ 
($\mh_{by}, \mh_{cay}, \mh_{uy}$)
versus metallicities estimated with the $hk_0,\ \vmy_0$ relation ($\mh_{uy}$).
The eight RGs with spectroscopic measurements from BA09 are marked as red stars.}
\end{figure}

\section{Metallicity distribution}
We select NGC~6522 RGs by star position as in Section~\ref{obs_cala}, 
in magnitude ($y <$ 18.0 mag), in photometric accuracy ($\sigma (v,b,y) <$ 0.03 
and $\sigma (Ca) <$ 0.02 mag), and in surface gravity 
($[c]=(u-v)-(v-b)-0.2\times(b-y) <$ 0.35 mag), ending up with 51 RGs.
A further selection is performed by using the proper motions 
by \citet{sumi} and imposing $-2 < pm(\alpha) < 6$ and $-2 < pm(\delta) < 2$ mas/yr 
(CA11). The final sample includes 28 candidate cluster RGs. 

Photometric metallicities are estimated by adopting the new theoretical 
calibration by CA11 based on the \strom $hk$ index. 
To unredden the $hk$ index we adopt $E(hk) = - 0.155\,E(\bmy)$ \citep{twa91}. 

The panels of Fig.~\ref{fig3} show the photometric metallicities of the 
28 candidate RGs estimated adopting the Metallicity--Index--Color relations 
based on $hk_0$ and the $\bmy,\ \camy,\ \umy$ unreddened colors plotted versus 
metallicities estimated with the $hk_0,\ \vmy$ relation. The eight RGs in 
common with BA09 are marked with a red star. 
Metallicity estimates based on the $hk_0, \, \bmy$ and on the $hk_0, \, \umy$ 
relations agree quite well with those based on the $hk_0, \, \vmy$ relation 
($<\Delta(\mh_{\hbox{\footnotesize by, uy}} - \mh_{\hbox{\footnotesize vy}})>\, \simeq 0.0\pm0.02$ and $\simeq -0.01\pm0.02$ 
with $\sigma$ = 0.26 and 0.19 dex, respectively.)
On the other hand, the metallicity estimates based on the $hk_0, \, \camy$ 
relation are on average $\approx$0.3 dex more metal-rich than those 
based on the $hk_0, \, \vmy$ relation. 
The difference might be due to the fact that the $hk_0, \, \camy$ relation is more 
sensitive to the $Ca$ abundance and, in turn, to the $\alpha$-element abundance,
than the other relations. 

\section{Conclusions and future perspectives}
We presented $Ca-uvby$ \strom photometry of the Baade's Window, 
including the GGC NGC~6522. 
We separated field and cluster stars and estimated the global metallicity of 
candidate RGs in NGC~6522 by using a new theoretical metallicity 
calibration of the $hk$ index.
We find that metallicities estimated by adopting the $hk$ index and the 
$\camy$ color are systematically  more metal-rich than metallicities 
estimated with the $hk$ index and the 
$\bmy, \vmy$ and $\bmy$ colors. The $hk$ index is then affected 
not only by the $Ca$ abundance, but also by another source
of opacity.

We now plan to perform a new $\alpha$-enhanced theoretical calibration of the 
\strom index $m_1$ up to solar metallicities, to constrain the metallicity
distribution of bulge RGs and red clump stars.
We then plan to match \strom data with $z,y,J,H,K$-band photometry 
of the bulge from the VVV public survey project 
(VISTA, ESO, http://vvvsurvey.org/).
The optical-NIR color-color planes will allow us to better disentangle the 
different components present in the observed 
fiel of view (thin/thick disk, bulge stars). 
We will then test the adoption of different intermediate- and 
broad-band color-color planes in order to study the metallicity and 
the age distribution of these stars.

\small  
%
%

%
%
%
%
%

\bibliographystyle{aa}

\bibliography{mnemonic,ref_calamida_2}

\end{document}